\documentclass{elsarticle}

\usepackage{lineno,hyperref}

\usepackage{amsmath}
\usepackage{amssymb}
\usepackage{amsthm}
\usepackage{txfonts}
\usepackage{dsfont}
\usepackage{amssymb}
\usepackage{amsfonts}
\usepackage{algorithm}
\usepackage{algorithmic}
\usepackage{graphicx}
\usepackage{float}

\modulolinenumbers[5]

\begin{document}

\newtheorem{theorem}{Theorem}[section]
\newtheorem{proposition}{Proposition}[section]
\newtheorem{lemma}{Lemma}[section]
\newtheorem{corollary}{Corollay}[section]
\newtheorem{example}{Example}[section]
\newtheorem{remark}{Remark}[section]
\newtheorem{definition}{Definition}[section]
\newcommand{\vect}[1]{\mathrm{vec}(#1)}

\renewcommand{\algorithmicrequire}{\textbf{Input:}}
\renewcommand{\algorithmicensure}{\textbf{Output:}}
\newcommand{\abs}[1]{\lvert#1\rvert}

\begin{frontmatter}

\title{Computing the determinant of a matrix with polynomial entries by approximation}

\author[SCU,Casit]{Xiaolin Qin\corref{cor1}}
\ead{qinxl@casit.ac.cn}
\author[Casit]{Zhi Sun\corref{cor1}}
\ead{sunzhi1019@163.com}
\author[Casit]{Tuo Leng}
\author[Casit]{Yong Feng}

\cortext[cor1]{Corresponding author}
\address[SCU]{Department of Mathematics, Sichuan University, Chengdu 610064, PR China}
\address[Casit]{Chengdu Institute of Computer Applications, Chinese Academy of Sciences, Chengdu 610041, PR China}

\begin{abstract}
Computing the determinant of a matrix with the univariate and multivariate polynomial entries arises frequently in the scientific computing and engineering fields. In this paper, an effective algorithm is presented for computing the determinant of a matrix with polynomial entries using hybrid symbolic and numerical computation. The algorithm relies on the Newton's interpolation method with error control for solving Vandermonde systems. It is also based on a novel approach for estimating the degree of variables, and the degree homomorphism method for dimension reduction. Furthermore, the parallelization of the method arises naturally.

\end{abstract}


\end{frontmatter}

\linenumbers

\section{Introduction}
In the scientific computing and engineering fields, such as
computing multipolynomial resultants \cite{Cox}, computing the
implicit equation of a rational plane algebraic curve given by its
parametric equations \cite{Delvaux2009}, and computing Jacobian
determinant in multi-domain unified modeling \cite{Qin2013}, computing the determinant of
a matrix with polynomial entries (also called symbolic determinant) is inevitable. Therefore,
computing symbolic determinants is an active area of research
[4--12]. There are several techniques for calculating the
determinants of matrices with polynomial entries, such as expansion
by minors \cite{Gentleman1976}, Gaussian elimination over the
integers \cite{Sasaki1982,Kaltofen1992}, a procedure which computes
the characteristic polynomial of the matrix \cite{Lipson1969}, and a
method based on evaluation and interpolation [5--7]. The first three algorithms belong to symbolic computations. As is well known, symbolic computations are
principally exact and stable. However, they have the disadvantage of
intermediate expression swell. The last one is the interpolation method,
which as an efficient numerical method has been widely used to
compute resultants and determinants, etc.. In fact, it is not
approximate numerical computations but big number computations,
which are also exact computations and only improve intermediate
expression swell problem. Nevertheless, the main idea of black box
approach takes an external view of a matrix, which is a linear
operator on a vector space \cite{CEK2002}. Therefore, it is
particularly suited to the handling of large sparse or structured
matrices over finite fields. In this paper, we propose an efficient
approximate interpolation approach to remedy these drawbacks.

Hybrid symbolic-numerical computation is a novel method for solving
large scale problems, which applies both numerical and symbolic
methods in its algorithms and provides a new perspective of them.
The approximate interpolation methods are still used to get the
approximate results [12--15]. In order to obtain exact results, one
usually uses exact interpolation methods to meliorate intermediate
expression swell problem arising from symbolic computations \cite{Marco2004,Li2009,Chen2013,Kaltofen2007}. Although the underlying floating-point methods in principle
allow for numerical approximations of arbitrary precision, the
computed results will never be exact. Recently, the exact
computation by intermediate of floating-point arithmetic has been an
active area of solving the problem of intermediate expression swell
in [16--20]. The nice feature of the work is as follows: The initial
status and final results are accurate, whereas the intermediate of
computation is approximate. The aim of this paper is to provide a
rigorous and efficient algorithm to compute symbolic determinants by
approximate interpolation. In this paper, we restrict our study to a
non-singular square matrix with polynomial entries and the
coefficients of polynomial over the integers.

The rest of this paper is organized as follows. Section 2 first
constructs the degree matrix of symbolic determinant on variables
and gives theoretical support to estimate the upper bounds degree of
variables, and then analyzes the error controlling for solving
Vandermonde systems of equations by Newton's interpolation method,
finally proposes a reducing dimension method based on degree
homomorphism. Section 3 proposes a novel approach for estimating the
upper bound on degree of variables in symbolic determinant, and
then presents algorithms of dimension reduction and lifting
variables and gives a detailed example. Section 4 gives some
experimental results. The final section makes conclusions.

\section{Preliminary results}
Throughout this paper, $\mathbb{Z}$ and $\mathbb{R}$ denote the set
of the integers and reals, respectively. There are $v$ variables
named $x_i$, for $i=1$ to $v$. Denote the highest degree of each
$x_i$ by $d_i$. Denoted by ${\Phi}_{m,n} (\mathbb{F})$ the set of all
$m$ by $n$ matrices over field $\mathbb{F} = \mathbb{R}$, and
abbreviate ${\Phi}_{n,n}(\mathbb{F})$ to ${\Phi}_{n}(\mathbb{F})$.

\subsection{Estimating degree of variables}
In this subsection, a brief description to Chio's expansion is
proposed. We also give the Theorem \ref{lem:maxdeg} for estimating the upper bound
on degree of variables in symbolic determinant.
\begin{lemma}\label{chio's theorem}
(\cite{Howard1966}) Let $A=[a_{ij}]$ be an $n\times n$ matrix and
suppose $a_{11}\neq 0$. Let $K$ denote the matrix obtained by
replacing each element $a_{ij}$ in $A$ by
$\begin{vmatrix}
a_{11}& a_{1j}\\
a_{i1}&a_{ij}
\end{vmatrix}$. Then $|A|=|K|/a_{11}^{n-2}$. That is,
\begin{displaymath}|A|=\frac{1}{a_{11}^{n-2}}
\begin{vmatrix}
\begin{vmatrix}
a_{11}&a_{12}\\
a_{21}&a_{22}
\end{vmatrix}
 &\begin{vmatrix}
a_{11}&a_{13}\\
a_{21}&a_{23}
\end{vmatrix}&\cdots &
\begin{vmatrix}
a_{11}&a_{1n}\\
a_{21}&a_{2n}
\end{vmatrix}\\
\begin{vmatrix}
a_{11}&a_{12}\\
a_{31}&a_{32}
\end{vmatrix}
 &\begin{vmatrix}
a_{11}&a_{13}\\
a_{31}&a_{33}
\end{vmatrix}&\cdots &
\begin{vmatrix}
a_{11}&a_{1n}\\
a_{31}&a_{3n}
\end{vmatrix}\\
\cdots &\cdots &\cdots &\cdots\\
\begin{vmatrix}
a_{11}&a_{12}\\
a_{n1}&a_{n2}
\end{vmatrix}
 &\begin{vmatrix}
a_{11}&a_{13}\\
a_{n1}&a_{n3}
\end{vmatrix}&\cdots &
\begin{vmatrix}
a_{11}&a_{1n}\\
a_{n1}&a_{nn}
\end{vmatrix}
\end{vmatrix}.
\end{displaymath}
\end{lemma}
\begin{remark}\label{remchio}
The proof of Lemma \ref{chio's theorem} is clear. Multiply each row
of $A$ by $a_{11}$ except the first, and then perform the elementary
row operations, denote $Op(2-a_{21}\cdot 1)$, $Op(3-a_{31}\cdot 1)$,
$\cdots$, $Op(n-a_{n1}\cdot 1)$, where $'1', '2', \cdots, 'n'$
represents for the row index. We get
\begin{displaymath}a_{11}^{n-1}|A|=
\begin{vmatrix}
a_{11}& a_{12}&\cdots &a_{1n}\\
a_{11}a_{21}& a_{11}a_{22}&\cdots &a_{11}a_{2n}\\
\vdots&\vdots&\ddots&\vdots\\
a_{11}a_{n1}& a_{11}a_{n2}&\cdots &a_{11}a_{nn}
\end{vmatrix}=
\end{displaymath}
\begin{displaymath}
\begin{vmatrix}
a_{11}& a_{12}&a_{13}&\cdots &a_{1n}\\
0& \begin{vmatrix}
a_{11}&a_{12}\\
a_{21}&a_{22}
\end{vmatrix}&\begin{vmatrix}
a_{11}&a_{13}\\
a_{21}&a_{23}
\end{vmatrix}&\cdots &\begin{vmatrix}
a_{11}&a_{1n}\\
a_{21}&a_{2n}
\end{vmatrix}\\
\vdots&\vdots&\vdots&\ddots&\vdots\\
0& \begin{vmatrix}
a_{11}&a_{12}\\
a_{n1}&a_{n2}
\end{vmatrix}&\begin{vmatrix}
a_{11}&a_{13}\\
a_{n1}&a_{n3}
\end{vmatrix}&\cdots &\begin{vmatrix}
a_{11}&a_{1n}\\
a_{n1}&a_{nn}
\end{vmatrix}
\end{vmatrix}
=a_{11}|K|.
\end{displaymath}
We observe that $K$ is $(n-1)\times (n-1)$ matrix, then the above procedure can be repeated until the $K$ is $2 \times 2$ matrix. It is a simple and straightforward method for calculating the determinant of a numerical matrix.
\end{remark}

\begin{lemma}\label{lem:deg}
Given two polynomials $f(x_1)$ and $g(x_1)$, the degree of the product
of two polynomials is the sum of their degrees, i.e.,
$$deg( f(x_1) \cdot g(x_1), x_1 ) = deg(f(x_1), x_1) + deg(g(x_1), x_1).$$
The degree of the sum (or difference) of two polynomials is equal to
or less than the greater of their degrees, i.e.,
$$deg(f(x_1) \pm g(x_1), x_1) \leq max\{deg(f(x_1), x_1),
deg(g(x_1), x_1)\}, $$ where $f(x_1)$ and $g(x_1)$ are the univariate
polynomials over field $\mathbb{F}$, and $deg(f(x_1), x_1)$ represents the
highest degree of $x_1$ in $f(x_1)$.
\end{lemma}

Let $M=[M_{ij}]$ be an $n\times n$ matrix and suppose $M_{ij}$ is a polynomial with integer coefficients consisting of variables $x_1, x_2, \cdots, x_v$, where the order of $M$ is $n\geq 2$. Without loss of generality, we call it the degree matrix $\Omega_1 = (\sigma_{ij})$ \footnote{$\Omega_1, \Omega_2, \cdots, \Omega_v$ denote the degree matrix of $x_1, x_2, \cdots, x_v$, respectively.} for $x_1$ defined as:
\begin{center}
${\sigma_{ij}}=\begin{cases}
 highest \ degree \ of \ x_1 \ appears\ in \ the \ element \ M_{ij}, i.e., deg(M_{ij}, x_1),\\
 0, \;\;\; if \ x_1 \ does \ not \ occur \ in \ M_{ij}.
\end{cases}$
\end{center}
So, we can construct the degree matrix from $M$ for all variables, respectively.

\begin{theorem}\label{lem:maxdeg}
$M$ is defined as above. Suppose the
 $2 \times 2$ degree matrix can be obtained from $M$ for $x_i (1\leq i \leq v)$,
denotes
\begin{eqnarray*}\Omega_i=
\left[\begin{array}{cc}
\sigma_{(n-1)(n-1)}^{(n-2)}&\ \ \ \ \sigma_{(n-1)n}^{(n-2)}\\
\sigma_{n(n-1)}^{(n-2)}&\ \ \sigma_{nn}^{(n-2)}\\
\end{array}\right],
\end{eqnarray*}
then
$$maxdeg=\max\{\sigma_{(n-1)(n-1)}^{(n-2)}+\sigma_{nn}^{(n-2)}, \sigma_{(n-1)n}^{(n-2)}+\sigma_{n(n-1)}^{(n-2)}\}.$$ That is,
the maximum degree of variable is no more than
$$maxdeg-\sum_{i=3}^{n}(i-2)\sigma_{(n-i+1)(n-i+1)}^{(n-i)},$$
where $\sigma^{(n-2)}_{(n-1)(n-1)}=deg(M_{(n-1)(n-1)}^{(n-2)}, x_i).$\footnote{$\sigma^{(\cdot)}_{ij}$ is defined by the same way for the rest of this paper.}
\end{theorem}
\begin{proof}
Considering the order $n$ of symbolic
determinant
\[
|M| = \left| {\begin{array}{cccc}
 M_{11}&\ M_{12}& \ \cdots &M_{1n}\\
M_{21}&\ M_{22}&\ \cdots &M_{2n}\\
\vdots&\ \vdots&\ddots&\ \vdots\\
M_{n1}&\ M_{n2}&\ \cdots &\ M_{nn}
 \end{array} } \right|
\]
by Chio's expansion is from Remark \ref{remchio}, then
\begin{displaymath}|M|=\frac{1}{M_{11}^{n-2}}
\left| {\begin{array}{cccc}
M_{22}^{(1)}&\ M_{23}^{(1)}&\ \cdots&\ M_{2n}^{(1)}\\
M_{32}^{(1)}&\ M_{33}^{(1)}&\ \cdots&\ M_{3n}^{(1)}\\
\vdots&\ \vdots&\ \ddots&\ \vdots\\
M_{n2}^{(1)}& \ M_{n3}^{(1)}&\ \cdots&\ M_{nn}^{(1)}\\
 \end{array} } \right|
\end{displaymath}
\begin{displaymath}
=\frac{1}{M_{11}^{n-2}}\frac{1}{{M_{22}^{(1)}}^{n-3}}\cdots\frac{1}{M_{(n-2)(n-2)}^{(n-3)}}\begin{vmatrix}
M_{(n-1)(n-1)}^{(n-2)}&\ M_{(n-1)n}^{(n-2)}\\
M_{n(n-1)}^{(n-2)}&\ M_{nn}^{(n-2)}
\end{vmatrix},
\end{displaymath}

where
$$M_{22}^{(1)}=M_{11}M_{22}-M_{12}M_{21}, M_{32}^{(1)}=M_{11}M_{32}-M_{12}M_{31}, \cdots, M_{nn}^{(1)}=M_{11}M_{nn}-M_{1n}M_{n1}.$$

By Lemma \ref{lem:deg}, for $x_i$ we get
$$
deg(|M|, x_i)\leq
\max\{\sigma_{(n-1)(n-1)}^{(n-2)}+\sigma_{nn}^{(n-2)},
\sigma_{(n-1)n}^{(n-2)}+\sigma_{n(n-1)}^{(n-2)}\} -(n-2)\sigma_{11}
-(n-3)\sigma_{22}^{(1)}-\cdots-\sigma^{(n-3)}_{(n-2)(n-2)}$$
$$= maxdeg-\sum_{i=3}^{n}(i-2)\sigma^{(n-i)}_{(n-i+1)(n-i+1)},$$
where
$$maxdeg=\max\{\sigma_{(n-1)(n-1)}^{(n-2)}+\sigma_{nn}^{(n-2)}, \sigma_{(n-1)n}^{(n-2)}+\sigma_{n(n-1)}^{(n-2)}\}.$$ The proof
of Theorem \ref{lem:maxdeg} is completed. It can be applied to all variables.
\end{proof}
\begin{remark}
We present a direct method for estimating the upper bound on degrees of variables by computation of the degree matrices. Our method only needs the simple recursive arithmetic operations of addition and subtraction. Generally, we can obtain the exact degrees of all variables in symbolic determinant in practice.
\end{remark}

\subsection{Newton's interpolation with error control}
Let $M$ be defined as above. Without loss of generality, we consider the determinant of a matrix with bivariate polynomial entries, and then generalize the results to the univariate or multivariate polynomial. A
good introduction to the theory of interpolation can be seen in
\cite{Boor1994}.

\begin{definition}The Kronecker product of ${A} = [a_{i,j}]\in
{\Phi}_{m,n}(\mathbb{F})$ and ${B} = [b_{ij}] \in
{\Phi}_{p,q}(\mathbb{F})$ is denoted by ${A}\otimes {B}$ and is defined
to the block matrix
\begin{equation}
{A}\otimes{B}=\left(\begin{array}{cccc}a_{11}{B}&a_{12}{B}&\cdots
&a_{1n}{B}\\
a_{21}{B}&a_{22}{B}&\cdots
&a_{2n}{B}\\
\vdots&\vdots&\ddots&\vdots\\
a_{m1}{B}&a_{m2}{B}&\cdots&a_{mn}{B}
\end{array}\right)\in{M}_{mp,nq}(\mathbb{F}).
\end{equation}
Notice that ${A}\otimes {B} \neq {B}\otimes {A} $ in general.
\end{definition}
\begin{definition}
With each matrix ${A} = [a_{ij}] \in {\Phi}_{m, n}(\mathbb{F})$, we
associate the vector $\vect{{A}}\in \mathbb{F}^{mn}$ defined by
$$\vect{{A}} \equiv [a_{11},\cdots
a_{m1},a_{12},\cdots,a_{m2},\cdots,a_{1n},\cdots,a_{mn}]^T,$$ where
$^T$ denotes the transpose of matrix or vector.
\end{definition}
Let the determinant of $M$ be
$f(x_1,x_2)=\sum_{i,j}a_{ij}x_1^ix_2^j$ which is a polynomial with
integer coefficients, and $d_1$, $d_2$ \footnote{$d_1, d_2$ are
defined by the same way for the rest of this paper.}be the
bounds on the highest degree of $f(x_1,x_2)$ in $x_1$, $x_2$,
respectively. We choose the distinct scalars $(x_{1i}, x_{2j})$ ($i=
0, 1, \cdots, d_1$; $j = 0, 1, \cdots, d_2$), and obtain the values
of $f(x_1, x_2)$, denoted by $f_{ij}\in\mathbb{R}$ ($i = 0, 1,
\cdots, d_1; j = 0, 1, \cdots, d_2$). The set of monomials is
ordered as follows:

$$(1, x_1, x_1^2, \cdots, x_1^{d_1}) \times (1, x_2, x_2^2, \cdots, x_2^{d_2}), $$

and the distinct scalars in the corresponding order is as follows:
$$
(x_{10}, x_{11}, \cdots, x_{1d_1}) \times (x_{20}, x_{21}, \cdots,
x_{2d_2}).
$$

Based on the bivariate interpolate polynomial technique, which is
essential to solve the following linear system:

\begin{equation}\label{equ:tem1}
({V}_{x_1}\otimes{V}_{x_2})\vect{{a}}=\vect{{F}},
\end{equation}

where the coefficients ${V}_{x_1}$ and ${V}_{x_2}$ are Vandermonde
matrices:
$${V}_{x_1}=\left(
\begin{array}{ccccc}
1&x_{10}&x_{10}^2&\cdots&x_{10}^{d_1}\\
1&x_{11}&x_{11}^2&\cdots&x_{11}^{d_1}\\
\vdots&\vdots&\vdots&\ddots&\vdots\\
1&x_{1d_1}&x_{1d_1}^2&\cdots&x_{d_1}^{1d_1}
\end{array}\right),\quad
{V}_{x_2}=\left(
\begin{array}{ccccc}
1&x_{20}&x_{20}^2&\cdots&x_{20}^{d_2}\\
1&x_{21}&x_{21}^2&\cdots&x_{21}^{d_2}\\
\vdots&\vdots&\vdots&\ddots&\vdots\\
1&x_{2d_2}&x_{2d_2}^2&\cdots&x_{2d_2}^{d_2}
\end{array}\right),
$$ and
$${a}=\left( \begin{array}{cccc}
a_{00}&a_{01}&\cdots& a_{0d_2}\\
a_{10}&a_{11}&\cdots& a_{1d_2}\\
\vdots&\vdots&\ddots&\vdots\\
a_{d_10}&a_{d_11}&\cdots&a_{d_1d_2}
\end{array}
 \right),\quad {F}=\left( \begin{array}{cccc}
f_{00}&f_{01}&\cdots& f_{0d_2}\\
f_{10}&f_{11}&\cdots& f_{1d_2}\\
\vdots&\vdots&\ddots&\vdots\\
f_{d_10}&f_{d_11}&\cdots&f_{d_1d_2}
\end{array}
 \right).$$

Marco et al. \cite{Marco2004} have proved in this way that the
interpolation problem has a unique solution. This means that
${V}_{x_1}$ and ${V}_{x_2}$ are nonsingular and therefore ${V}=
{V}_{x_1}\otimes{V}_{x_2}$, then the coefficient matrix of the linear
system (\ref{equ:tem1}) is nonsingular. The following lemma shows
us how to solve the system (\ref{equ:tem1}).
\begin{lemma}\label{theo:kronecker_equation}
(\cite{Horn1991}) Let $\mathbb{F}$ denote a field. Matrices
${A}\in{\Phi}_{m,n}(\mathbb{F})$, ${B}\in{\Phi}_{q,p}(\mathbb{F})$, and
${C}\in{\Phi}_{m,q}(\mathbb{F})$ are given and assume
${X}\in{\Phi}_{n,p}(\mathbb{F})$ to be unknown. Then, the following
equation:
\begin{equation}\label{equ:kronecker_equ}
({B}\otimes{A})\vect{{X}}=\vect{{C}}
\end{equation}
is equivalent to  matrix equation:
\begin{equation}\label{equ:matrix_equ}
{AXB}^T={C}.
\end{equation}
Obviously, equation (\ref{equ:matrix_equ}) is equivalent to the
system of equations
\begin{equation}\left\{\begin{array}{l}
{AY}={C} \\
{BX}^T={Y}^T.
\end{array}\right.
\end{equation}
\end{lemma}
Notice that the coefficients of system (\ref{equ:tem1}) are
Vandermonde matrices, the reference \cite{Bjorck1970} by the
Newton's interpolation method presented a progressive algorithm
which is significantly more efficient than previous available
methods in $O(d_1^2)$ arithmetic operations in Algorithm
\ref{alg:dual}.
\newcounter{num}
\begin{algorithm}[H]
\caption{(Bj\"{o}rck and Pereyra algorithm)} \label{alg:dual}
Input: a set of distinct scalars $(x_i, f_i) (0\leq i \leq d_1)$;\\
Output: the solution of coefficients $a_0, a_1, \cdots, a_{d_1}$.
\begin{list}{Step \arabic{num}:}{\usecounter{num}\setlength{\rightmargin}{\leftmargin}}
\item
\begin{algorithmic}
  $c_i^{(0)} := f_i(i=0, 1, \cdots, d_1)$\\
    \FOR {$k= 0$ to $d_1-1$}
         \STATE
          $c_{i}^{(k+1)} := \frac{c_i^{(k)}-c_{i-1}^{(k)}}{x_i-x_{i-k-1}}(i=d_1, d_1-1, \cdots,
         k+1)$
    \ENDFOR
\end{algorithmic}
\item
\begin{algorithmic}
$a_i^{(d_1)} := c_i^{(d_1)}(i=0, 1, \cdots, d_1)$\\
 \FOR {$k= d_1-1$ to $0$ by $-1$}
           \STATE
                      $a_{i}^{(k)} := a_i^{(k+1)}-x_{k}a_{i+1}^{(k+1)}(i=k, k+1, \cdots,
         d_1-1)$
         \ENDFOR
\end{algorithmic}
\item
Return $a_i := a_i^{(0)}(i=0, 1, \cdots, d_1)$.
\end{list}
\end{algorithm}

In general, we can compute the equation (\ref{equ:tem1}) after
choosing $d_1+1$ distinct scalars $(x_{10},x_{11},\cdots,x_{1d_1})$
and $d_2+1$ distinct scalars $(x_{20}, x_{21}, \cdots, x_{2d_2})$,
then obtain their corresponding exact values $(f_{00},f_{01},
\cdots,f_{0d_2},\cdots, $\\ $f_{10},f_{11},\cdots,f_{1d_2},\cdots,f_{d_10},f_{d_11},\cdots, f_{d_1d_2})$. However, in order to improve
intermediate expression swell problem arising from symbolic
computations and avoid big integer computation, we can get the
approximate values of $f(x_1, x_2)$, denoted by $(\tilde{f}_{00},
\tilde{f}_{01}, $\ $ \cdots, \tilde{f}_{0d_2},\tilde{f}_{10},
\tilde{f}_{11}, \cdots, \tilde{f}_{1d_2}, \tilde{f}_{d_10},
\tilde{f}_{d_11},\cdots,$\ $\tilde{f}_{d_1d_2})$.

Based on Algorithm \ref{alg:dual}, together with Lemma
\ref{theo:kronecker_equation} we can obtain the approximate solution
$\tilde{{a}}=[\tilde{a}_{ij}]$($i = 0, 1, \cdots, d_1; j = 0, 1,
\cdots, d_2$). So an approximate bivariate polynomial
$\tilde{f}(x_1, x_2) = \sum_{i,j}\tilde{a}_{ij}x_1^ix_2^j$ is only
produced. However, we usually need the exact results in practice.
Next, our main task is to bound the error between approximate
coefficients and exact values, and discuss the controlling error
$\varepsilon$ in Algorithm \ref{alg:dual}. The literature
\cite{Feng2011} gave a preliminary study of this problem. Here, we
present a necessary condition on error controlling $\varepsilon$ in
floating-point arithmetic. In Step 1 of Algorithm \ref{alg:dual}, it
is the standard method for evaluating divided
differences($c_k^{(k)}=f[x_0, x_1, \cdots, x_k]$). We consider the
relation on the $f_{ij}-\tilde{f}_{ij}$ with $a_{ij}-\tilde{a}_{ij}$
and the propagation of rounding errors in divided difference
schemes. We have the following theorem to answer the above question.
\begin{lemma}\label{lem:unierror}
$c_i$ and $f_{i}$ are defined as in Algorithm \ref{alg:dual},
$\tilde{c}_i$ and $\tilde{f}_{i}$ are their approximate values by
approximate interpolation, $\lambda=\min\{|x_{2i}-x_{2j}|: i \neq
j\}(0<\lambda<1)$. Then
$$|c_i- \tilde{c}_i|\leq
(\frac{2}{\lambda})^{d_2}\max\{|f_{i}-\tilde{f}_{i}|\}.$$
\end{lemma}
\begin{proof}
From Algorithm \ref{alg:dual}, we observe that Step 1 is recurrences
for $c_i^{(k+1)}, (k=0, 1, \cdots, d_2-1, i=d_2, d_2-1, \cdots,
k+1)$, whose form is as follows:
\begin{eqnarray*}
c_i^{(d_2)}=\frac{1}{\lambda}(c_i^{(d_2-1)}-c_{i-1}^{(d_2-1)}).
\end{eqnarray*}
However, when we operate the floating-point arithmetic in Algorithm
\ref{alg:dual}, which is recurrences for $\tilde{c}_i^{(k+1)}$,
which form is as follows:
\begin{eqnarray*}
\tilde{c}_i^{(d_2)}=\frac{1}{\lambda}(\tilde{c}_i^{(d_2-1)}-\tilde{c}_{i-1}^{(d_2-1)}).
\end{eqnarray*}
Therefore,
\begin{eqnarray*}
|c_i^{(d_2)}-
\tilde{c}_i^{(d_2)}|=\frac{1}{\lambda}|c_i^{(d_2-1)}-\tilde{c}_i^{(d_2-1)}+\tilde{c}_{i-1}^{(d_2-1)}-c_{i-1}^{(d_2-1)}|
\leq\frac{1}{\lambda}(|c_i^{(d_2-1)}-\tilde{c}_i^{(d_2-1)}|+|c_{i-1}^{(d_2-1)}-\tilde{c}_{i-1}^{(d_2-1)}|).
\end{eqnarray*}
The bounds are defined by the following recurrences,
\begin{eqnarray*}
|c_i^{(d_2)}-
\tilde{c}_i^{(d_2)}|\leq\frac{2}{\lambda}|c_{i-1}^{(d_2-1)}-\tilde{c}_{i-1}^{(d_2-1)}|\leq\cdots\leq
(\frac{2}{\lambda})^{d_2}\max\{|f_{i}-\tilde{f}_{i}|\}.
\end{eqnarray*}
This completes the proof of the lemma.
\end{proof}

\begin{theorem}\label{theo:errocontrol} Let
$\varepsilon=\max\{|f_{ij}-\tilde{f}_{ij}|\}$,
$\lambda=\min\{|x_{1i}-x_{1j}|, |x_{2i}-x_{2j}|: i \neq
j\}(0<\lambda<1)$. Then
$$\max\{|a_{ij}-\tilde{a}_{ij}|\}\le(\frac{2}{\lambda})^{d_1}(\frac{2}{\lambda})^{d_2}\varepsilon.
$$
\end{theorem}
\begin{proof}
From equation (\ref{equ:tem1}), it holds that
$${V}\vect{{\tilde{a}}-{a}}=\vect{{\tilde{F}}-{F}},$$
where ${V} = {V}_{x_1} \otimes {V}_{x_2}$. By Lemma
\ref{theo:kronecker_equation}, the above equation is equivalent to
the following equation:
$${V}_{x_2}{({\tilde{a}}-{a})}{V}_{x_1}^T={\tilde{F}}-{F}.$$
Thus, it is equivalent to
\begin{subequations}
\begin{eqnarray}
&&{V}_{x_2}{z}={\tilde{F}}-{F} \label{equ:a}\\
&&{V}_{x_1}({\tilde{a}}-{a})^T={z}^T \label{equ:b}
\end{eqnarray}
\end{subequations}
where ${z}=[z_{ij}]$. Matrix equation (\ref{equ:a}) is equivalent to
\begin{equation}
{V}_{x_2}{z}_{.i}={\tilde{F}}_{i.}-{F}_{i.}, \quad i=1, 2, \cdots d_2+1
\end{equation}
where ${z}_{.i}$ stands for the $i$-th column of ${z}$ and
${F}_{i.}$ the $i$-th row of matrix ${F}$.

From Lemma \ref{lem:unierror} and Algorithm \ref{alg:dual}, it holds that
$$\max_{j=0}^{d_2}|z_{ji}|<(\frac{2}{\lambda})^{d_2}|f_{i\cdot}-\tilde{f}_{i\cdot
}|, \ for \ each \ i.$$ Hence, we conclude that
$$\max_{i,j}|z_{ji}|<(\frac{2}{\lambda})^{d_2}|f_{i\cdot}-\tilde{f}_{i\cdot
}|.$$

Let $\delta =(\frac{2}{\lambda})^{d_2}|f_{i\cdot}-\tilde{f}_{i\cdot
}|$, argue equation (\ref{equ:b}) in the same technique as do above,
we deduce that
$$\max_{i,j}|a_{ij}-\tilde{a}_{ij}|\leq (\frac{2}{\lambda})^{d_1}(\frac{2}{\lambda})^{d_2}\varepsilon. $$ The proof is
finished.
\end{proof}

In order to avoid the difficulty of computations, we restrict our
study to the coefficients of polynomial over $\mathbb{Z}$. So we need to solve the
Vandermonde system and take the nearest integer to each component of
the solution. The less degree of bounds on variables we obtain, the
less the amount of computation is for obtaining approximate
multivariate polynomial. Once an upper bound $d_1$ and $d_2$ are
gotten, we choose $(d_1+1)\cdot(d_2+1)$ interpolate nodes and
calculate
\begin{equation}\label{equ:error_control}
\varepsilon=0.5{(\frac{\lambda}{2})}^{d_1+d_2}.
\end{equation}
Then, compute the values $\tilde{f}_{ij}\approx f(x_{1i},x_{2j})$
for $i=0,1,\cdots,d_1,$\ $ j=0,1,\cdots,d_2$ with an error less than
$\varepsilon$. By interpolation method, we compute the approximate
interpolation polynomial $\tilde{f}(x_1, x_2)$ with coefficient
error less than 0.5.

As for the generalization of the algorithm to the case $v>2$, we can
say that the situation is completely analogous to the bivariate
case. It comes down to solving the following system:
\begin{equation}\label{equ:general}
\underbrace{({V}_{x_1}\otimes{V}_{x_2}\cdots\otimes{V}_{x_v})}_{v}\vect{{a}}
= \vect{{F}}.
\end{equation}
Of course, we can reduce the multivariate polynomial entries to
bivariate ones on symbolic determinant. For more details refer to
Section 2.3.

We can analyze the computational complexity of the derivation of
above algorithm. For the analysis of floating-point arithmetic operations,
the result is similar with the exact interpolation situation
\cite{Marco2004}. However, our method can enable the
practical processing of symbolic computations in applications.

\begin{remark}
Our result is superior to the literature \cite{Feng2011}. Here we
make full use of advantage of arbitrary precision of floating-point
arithmetic operations on modern computer and symbolic computation
platform, such as Maple. In general, it seems as if at least
some problems connected with Vandermonde systems, which
traditionally have been considered too ill-conditioned to be
attached, actually can be solved with good precision.
\end{remark}

\subsection{Reducing dimension method}
As the variables increased, the storage of computations expands
severely when calculated high order on symbolic determinant. The
literature \cite{Moenck1976} is to map the multivariate problem into
a univariate one. For the general case, the validity of the method is
established by the following lemma.
\begin{lemma}\label{lem:reducedim}
(\cite{Moenck1976}) In the polynomial ring $R[x_1, x_2, \cdots,
x_v], v>2$. The mapping:
\begin{eqnarray*}
\phi: R[x_1, x_2, \cdots, x_v] \rightarrow R[x_1] \\
\phi: x_i \mapsto x_1^{n_i}, 1\leq i \leq v
\end{eqnarray*}
where $n_v> n_{v-1}>\cdots >n_1=1$ is a homomorphism of rings.
\end{lemma}
Let $d_i(f(x_1, x_2, \cdots, x_v))$ be the highest degree of the
polynomial $f(x_1, x_2, \cdots, x_v)$ in variable $x_i$. The
following lemma relates the ${n_i}$ of the mapping to $d_i$ and
establishes the validity of the inverse mapping.
\begin{lemma}\label{lem:liftingvar}
(\cite{Moenck1976}) Let $\psi$ be the homomorphism of free R-modules defined by:\\
\begin{center}
$\psi: R[x_1] \rightarrow R[x_1, x_2, \cdots, x_v]$\\
$\psi: x_1^{k} \mapsto \makeatletter
\begin{cases}
         1             \ \ \ \  \ \ \ \ \ \ \ \ \ \ \ \ \ \ \ \ \ \ \ \ \ \ \ \ \mbox{if~ $k=0,$} \\
         \psi(x_1^r)\cdot x_i^q  \ \ \ \ \ \ \ \ \ \ \ \ \ \ \ \mbox{otherwise} \\
        \end{cases}
$
\end{center} where $n_{i+1}>k\geq n_i, k= q\cdot n_i+ r, 0\leq r <
n_i$ and $
n_v>\cdots >n_1 =1$.\\
Then for all $f(x_1, x_2, \cdots, x_v)\in R[x_1, x_2, \cdots, x_v],
\psi(\phi(f)) = f$, and for all $i$ if and only if
\begin{equation}\label{equ:condition}
\sum_{j=1}^{i}d_j(f)n_j<n_{i+1}, 1 \leq i< v.
\end{equation}
\end{lemma}
\begin{remark}
We apply the degree homomorphism method to reduce dimension for
computing the determinant of a matrix with multivariate polynomial entries, which is distinguished from the practical fast
polynomial multiplication \cite{Moenck1976}. We note that relation
(\ref{equ:condition}) satisfying is isomorphic to their univariate
images. Thus any polynomial ring operation on entries of symbolic
determinant, giving results in the determinant, will be preserved
by the isomorphism. In this sense $\phi$ behaves like a ring
isomorphism on the symbolic determinant of polynomials. Another way
to view the mapping given in the theorems is:
$$\phi: x_i \mapsto x_{i-1}^{n_i}, 2\leq i \leq v.$$
\end{remark}

\section{Derivation of the algorithm}
The aim of this section is to describe a novel algorithm for
estimating the degree of variables on symbolic determinant, and the
degree homomorphism method for dimension reduction.
\subsection{Description of algorithm}
Algorithm \ref{alg:maxdeg} is to estimate the degree of variables on
symbolic determinant by computation of the degree matrix, and
Algorithm \ref{alg:reducedim} and \ref{alg:liftingvar} are used to
reduce dimension and lift variables.
\begin{algorithm}[ht]
\caption{(Estimating degree of variables
algorithm)}\label{alg:maxdeg}
Input: given the order $n$ of symbolic determinant $M$, list of variables $vars$;\\
Output: the exact or upper bounds on degree of variables.
\begin{list}{Step \arabic{num}:}{\usecounter{num}\setlength{\rightmargin}{\leftmargin}}
\item Select variable from $vars$ respectively, and repeat the following steps
\begin{algorithmic}[1]
\LOOP
 \STATE Obtain the degree matrix $\Omega=(\sigma_{ij}) (1 \leq i,j\leq n)$ from $M$;
 \IF {order($\Omega$)=2}
 \STATE $maxdeg := \max\{\sigma_{11} + \sigma_{22}, \sigma_{12} + \sigma_{21}\}$
 \ELSE
   \FOR {$i= 1$ to $n-1$}
          \FOR {$j=1$ to $n-1$}
           \STATE $temp := \sigma_{i1}+\sigma_{1j}$
           \STATE $\sigma_{ij} := \max\{\sigma_{ij}+\sigma_{11}, temp\}$
   \ENDFOR
    \ENDFOR
    \ENDIF
 \FOR {$i=1$ to $n-2$}
       \STATE   $maxdeg := maxdeg - \sigma_{11}$
          \ENDFOR
\item Return $maxdeg$
\ENDLOOP
\end{algorithmic}
\end{list}
\end{algorithm}
\begin{theorem}
Algorithm \ref{alg:maxdeg} works correctly as specified and its
complexity is $O(n^2)$, where $n$ is the order of symbolic
determinant.
\end{theorem}
\begin{proof}
 Correctness of the algorithm follows from Theorem \ref{lem:maxdeg}.\\
 The number of arithmetic operations required to execute $(n-1)\times(n-1)$ additions and simultaneous
 comparisons, and remain $n-2$ substructions and one comparison by using degree
 matrix. Therefore, the total arithmetic operations are $n^2-n$,
 that is $O(n^2)$.
 \end{proof}
\begin{algorithm}[H]
\caption{(Reducing dimension algorithm)} \label{alg:reducedim}
Input: given the order $n$ of symbolic determinant $M$, list of variables $vars$;\\
Output: the order $n$ of symbolic determinant $M'$ with bivariate
polynomial entries.
\begin{list}{Step \arabic{num}:}{\usecounter{num}\setlength{\rightmargin}{\leftmargin}}
\item Call Algorithm \ref{alg:maxdeg} to evaluate the bounds on degree of the variables in $M$, denoted by $d_i(1\leq i\leq v)$.
\item Reducing dimension
\begin{algorithmic}[1]
 \STATE Divide the $vars$ into the partitions: $[x_1, x_2, \cdots, x_t], [x_{t+1}, x_{t+2}, \cdots,
 x_v]$;
\FOR{$i=t-1$ to $1$ by $-1$}
 \STATE $D_i := \prod_{j=i+1}^{t}(d_j+1)$, $x_i \leftarrow x_t^{D_i}$
 \ENDFOR
\FOR{$i=v-1$ to $t+1$ by $-1$}
 \STATE $D_i := \prod_{j=i+1}^{v}(d_j+1)$, $x_i\leftarrow x_v^{D_i}$
 \ENDFOR
\end{algorithmic}
\item Obtain the symbolic determinant $M'$ on variables $vars=[x_t, x_v$];
\item Return $M'$.
\end{list}
\end{algorithm}

\begin{remark}
The beauty of this method is in a substitution trick. In Algorithm
\ref{alg:reducedim}, $t = ceil(\frac{n}{2})$, where $ceil(c)$ is a
function which returns the smallest integer greater than or equal
the number $c$. We note that the lexicographic order $x_v \succ
x_{v-1}\succ \cdots \succ x_1$ and divide the $vars$ into two parts.
Then the symbolic determinant can be translated
into the entries with bivariate polynomial. It can be highly
parallel computation when the variables are more than three.
\end{remark}

\begin{algorithm}[ht]
\caption{(Lifting variables algorithm)}\label{alg:liftingvar}
Input: given the set of monomial on $x_t, x_v$ in $L$;\\
Output: the polynomial with $x_1, x_2, \cdots, x_v$.
\begin{list}{Step \arabic{num}:}{\usecounter{num}\setlength{\rightmargin}{\leftmargin}}
\item Obtain the corresponding power set on $x_t, x_v$, respectively;
\item Lifting variables
\begin{algorithmic}[1]
 \STATE Call Algorithm \ref{alg:reducedim}, extract the power $D_i(1\leq i \leq t-1, t+1 \leq i \leq v-1)$;
\WHILE{nops(L)$\neq$ NULL}
 \STATE $temp := deg(x_t)$
 \FOR{$i=1$ to $t-1$ by $1$}
 \STATE $d_i := iquo(temp, D_i), temp := irem(temp, D_i)$
 \ENDFOR
 \STATE $d_i := temp, temp := deg(x_v)$
 \FOR{$i=t+1$ to $v-1$ by $1$}
     \STATE $d_i := iquo(temp, D_i), temp := irem(temp, D_i)$
 \ENDFOR
 \STATE $d_i := temp$
\ENDWHILE
\end{algorithmic}
\item Obtain the new set of monomial $L'$ on $x_1, x_2, \cdots,
x_v$;
\item Return $L'$.
\end{list}
\end{algorithm}
\begin{remark}
To sum up, based on Algorithm \ref{alg:maxdeg} to estimate bounds on
degree of variables, Algorithm \ref{alg:reducedim} to reduce
dimension for multivariate case, Algorithm \ref{alg:dual} to solve the
Vandermonde coefficient matrix of linear equations with error
controlling, and finally Algorithm \ref{alg:liftingvar} to lift
variables for recovering the multivariate polynomial.

In this paper, we consider the general symbolic determinant, which
is not sparse. Applying the substitutions to the matrix entries as
described above and assuming the monomial exists in the determinant
then the bivariate form of unknown polynomial is a highest degree of
\begin{equation}
D=\sum_{i=1}^{ceil(\frac{n}{
2})}(d_i\cdot\prod_{k=i+1}^{ceil(\frac{n}{ 2})}(d_k+1)).
\end{equation}
While this upper bound on degree of variable is often much larger than needed, which is the
worst case and thus is suitable to all cases.
\end{remark}

\subsection{A small example in detail}

\begin{example}\label{exam1} For convenience and space-saving
purposes, we choose the symbolic determinant is three variables and
order 2 as follows.
\begin{displaymath}|M|=
\begin{vmatrix}
5x_1^2-3x_1x_2+2x_3^2&\ \ \ \ \ \ \ -9x_1-3x_2^2-x_3^2\\
-x_1+x_2+3x_2x_3 &x_3-4x_2^2
\end{vmatrix},
\end{displaymath}
At first, based on Algorithm \ref{alg:maxdeg} we estimate the degree
on $x_1, x_2, x_3$. For the variable $x_1$, we get
\begin{eqnarray*}\Omega_1=
\left[\begin{array}{cc}
2&\ \ 1\\
1&\ \ 0
\end{array}\right].
\end{eqnarray*}
Then $$\max\{2+0,1+1\} = 2.$$ Therefore, the maximum degree of the variable
$x_1$ is $2$. As the same technique for $x_2, x_3$, we can get $3$ and $3$.

Call Algorithm \ref{alg:reducedim}, by substituting $x_1=x_2^4$, we get
\begin{displaymath}|M'|=
\begin{vmatrix}
5x_2^8-3x_2^5+2x_3^2&\ \ \ \ \ \ \ -9x_2^4-3x_2^2-x_3^2\\
-x_2^4+x_2+3x_2x_3&x_3-4x_2^2
\end{vmatrix}.
\end{displaymath}
Then, based on Algorithm \ref{alg:maxdeg} we again estimate the
degree on $x_2, x_3$ for $[10, 3]$.

Based on the derivation of algorithm in Section 3.1 and Algorithm
\ref{alg:dual}, computing exact polynomial $f(x_2, x_3)$ as follows: Choose the
different floating-point interpolation nodes by using the distance
between two points 0.5; $\lambda=0.5$, compute
$\varepsilon=0.745\times10^{-8}$ from Theorem \ref{theo:errocontrol}. Compute the approximate
interpolate datum $\tilde{f}_{ij}$ such that
$|f_{ij}-\tilde{f}_{ij}|<\varepsilon$. We get the following
approximate bivariate polynomial:
$$4.99995826234x_2^8x_3-20.0000018736x_2^{10}+24.0010598569x_2^5x_3+12.0025760656x_2^7+2.00000000000x_3^3$$ $$-8.00094828634x_2^2x_3^2
-9.00045331720x_2^8+9.01977448800x_2^5-3.00897542075x_2^6
+3.02270681750x_2^3$$ $$+9.00076124850x_2^3x_3-1.00207248277x_2^4x_3^2
+1.00018098282x_2x_3^2+2.99986559933x_2x_3^3.$$
 Next, based on Algorithm
\ref{alg:liftingvar} we lift the variables to obtain the following
multivariate polynomial:
$$
4.99995826234x_1^2x_3-20.0000018736x_2^2x_1^2+24.0010598569x_1x_2x_3
+12.0025760656x_2^3x_1+2.00000000000x_3^3$$ $$-8.00094828634x_2^2x_3^2
-9.00045331720x_1^2+9.01977448800x_1x_2-3.00897542075x_2^2x_1
+3.02270681750x_2^3$$ $$+9.00076124850x_2^3x_3-1.00207248277x_1x_3^2+1.00018098282x_2x_3^2+2.99986559933x_2x_3^3.$$
Finally, we easily recover the integer coefficients of above approximate polynomial to the nearest values as follows:
$$
5x_1^2x_3-20x_1^2x_2^2+24x_1x_2x_3+12x_1x_2^3+2x_3^3-8x_3^2x_2^2
-9x_1^2+9x_1x_2-3x_2^2x_1+3x_2^3+9x_2^3x_3-x_3^2x_1+x_3^2x_2+3x_3^3x_2.$$
\end{example}

\section{Experimental results}
Our algorithms are implemented in \emph{Maple}. The following
examples run in the same platform of \emph{Maple } under Windows
and \textsc{amd} Athlon(tm) 2.70 Ghz, 2.00 GB of main memory(RAM).
Figures 1 and 2 present the $Time$ and $RAM$ of computing for
symbolic determinants to compare our method with symbolic
method($det$, see \emph{Maple}'s help), and exact interpolation
method \cite{Marco2004,Li2009,Chen2013}. Figure 1 compared with time for
computing, Figure 2  compared with memory consumption for
computing, the $order$ of $x$-coordinate represents for the order of
symbolic determinants.

\begin{figure}[!ht]
\centering \resizebox*{7cm}{!}{\includegraphics{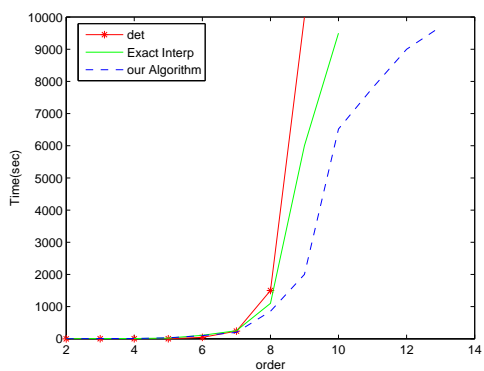}}
\caption{Computing time for symbolic determinant with different
algorithms}\label{Time}
\end{figure}

\begin{figure}[!ht]
\centering \resizebox*{7cm}{!}{\includegraphics{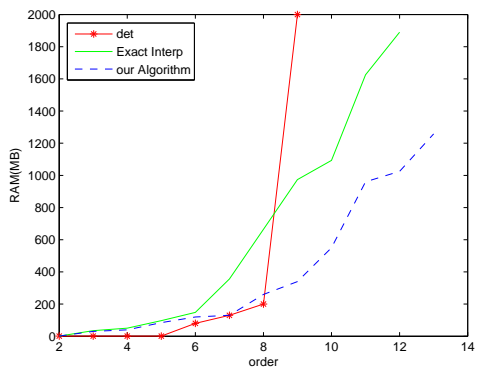}}
\caption{Computing memory for symbolic determinant with different
algorithms}\label{Memory}
\end{figure}

From Figures 1 and 2, we have the observations as follows:

\begin{enumerate}
 \item In general, the $Time$ and $RAM$ of algorithm $det$ are reasonable
when the $order$ is less than nine, and two indicators increase very
rapidly when the $order$ is to nine. However, two indicators of
interpolation algorithm is steady growth.
 \item Compared with the exact interpolation method, the approximate interpolation algorithm has the obvious advantages on the $Time$ and $RAM$
when the $order$ is more than eight.
\end{enumerate}
\begin{remark}
All examples are randomly generated using the command of
\textit{Maple}. The symbolic method has the advantage of the low
order or sparse symbolic determinants, such as expansion by minors,
Gaussian elimination over the integers. However, a purely symbolic
algorithm is powerless for many scientific computing problems, such
as resultants computing, Jacobian determinants and some practical
engineering always involving high-order symbolic determinants.
Therefore, it is necessary to introduce numerical methods to improve
intermediate expression swell problem arising from symbolic
computations.
\end{remark}

\section{Conclusions}
In this paper, we propose a hybrid
symbolic-numerical method to compute the symbolic determinants.
Meanwhile, we also present a novel approach for estimating the
bounds on degree of variables by the extended numerical determinant
technique, and introduce the reducing dimension algorithm. Combined
with these methods, our algorithm is more efficient than exact
interpolation algorithm for computing the high order symbolic determinants. It can be applied in
scientific computing and engineering fields, such as computing
Jacobian determinants in particular. Thus we can take fully
advantage of approximate methods to solve large scale symbolic
computation problems.

\bibliography{mybibfile}

\end{document}